\documentclass[journal]{IEEEtran}
\makeatletter
\usepackage{bm}
\usepackage{amssymb}
\usepackage{amsmath}
\usepackage{amsfonts}
\usepackage{graphicx}
\usepackage{subfigure}
\usepackage{cite}
\usepackage{enumerate}
\usepackage{url}
\usepackage{makecell} 
\usepackage{color}

\usepackage{algorithmic}
\usepackage{algorithm} 
\newtheorem{definition}{\bf Definition}
\newtheorem{proposition}{\bf Proposition}
\newtheorem{theorem}{\bf Theorem}

\newcommand{\beq}{\begin{equation}}
\newcommand{\eeq}{\end{equation}}
\newcommand{\beqq}{\begin{equation*}}
\newcommand{\eeqq}{\end{equation*}}

\DeclareMathOperator*{\argmax}{argmax}

\begin{document}

\title{\huge{Reinforcement Learning for Decentralized Trajectory Design in Cellular UAV Networks with Sense-and-Send Protocol}}
\author{Jingzhi Hu, \IEEEmembership{Student~Member,~IEEE},
Hongliang Zhang, \IEEEmembership{Student~Member,~IEEE},\\
and Lingyang Song, \IEEEmembership{Senior~Member,~IEEE}% <-this % stops a space

\thanks{
The authors are with School of Electrical Engineering and Computer Science, Peking University, Beijing, China (email: \{jingzhi.hu, hongliang.zhang, lingyang.song\}@pku.edu.cn).}
% \thanks{
% Y. Liao is with Department of Electrical and Computer Engineering, University of California, San Diego, the United States of America (email: yunliao@ucsd.edu).}%\thanks{
%	T. Wang is with School of Electronic Science and Engineering, Nanjing University, Nanjing, China (email: tianyu.alex.wang@nju.edu.cn).}
}

\maketitle

\begin{abstract}
Recently, the unmanned aerial vehicles (UAVs) have been widely used in real-time sensing applications over cellular networks, which sense the conditions of the tasks and transmit the real-time sensory data to the base station (BS). 
\textcolor[rgb]{0,0,0}{
The performance of a UAV is determined by the performance of both its sensing and transmission processes, which are influenced by the trajectory of the UAV. 
However, it is challenging for UAVs to design their trajectories efficiently, since they work in a dynamic environment.}
To tackle this challenge, in this paper, we adopt the reinforcement learning framework to solve the UAV trajectory design problem in a decentralized manner. 
To coordinate multiple UAVs performing the real-time sensing tasks, we first propose a sense-and-send protocol, and analyze the probability for successful valid data transmission using nested Markov chains. 
Then, we formulate the decentralized trajectory design problem and propose an enhanced multi-UAV Q-learning algorithm to solve this problem. 
Simulation results show that the proposed enhanced multi-UAV Q-learning algorithm converges faster and achieves higher utilities for the UAVs in the real-time task-sensing scenarios.
\end{abstract}


%%%%%%%%%%%%%%%%%%%%%%%
\section{Introduction}%
%%%%%%%%%%%%%%%%%%%%%%%
In the upcoming 5G network, the use of UAVs to perform sensing has been of particular interests, due to their high mobility, flexible deployment, and low operational cost~\cite{wang2017taking}. 
Specially, the UAVs have been wildly applied to execute critical sensing missions, such as traffic monitoring \cite{puri2007statistical}, precision agriculture \cite{alsalam2017autonomous}, and forest fire surveillance \cite{casbeer2006cooperative}. 
In these UAV sensing applications, the sensory data collected by the UAVs needs to be transmitted to the base station (BS) immediately for further real-time data processing. 
This poses a significant challenge for the UAVs to sense the task and send the collected sensory data simultaneously with a satisfactory performance.

In order to enable the real-time sensing applications, the cellular network controlled UAV transmission is considered as one promising solution \cite{van2016lte,zhang2018cellular2}, in which the uplink QoS is guaranteed compared to that in ad-hoc sensing networks \cite{thammawichai2018optimizing}. 
However, it remains a challenge for the UAVs to determine their trajectories in such cellular UAV networks. 
When the UAV is far from the task, it risks in obtaining invalid sensing data, while if it is far from the BS, the low uplink transmission quality may lead to difficulties in transmitting the sensory data to the BS. 
Therefore, the UAVs need to take both the sensing accuracy and the uplink transmission quality into consideration in designing their trajectories. 
Moreover, it is even more challenging when the UAVs belong to different entities and are uncooperative. 
Since the spectrum resource is scarce, the UAVs performing different sensing tasks have the incentive to compete for the limited uplink channel resources. In this regard, the UAVs have to consider the movement of other UAVs, which makes them work in a dynamic environment.
Therefore, a decentralized trajectory design approach is necessary for the UAVs real-time sensing problem, in which the location of the task and the BS and the behaviors of the other UAVs have to be taken in to consideration. 

To tackle these challenges, in this paper, we adopt the reinforcement learning framework to solve the UAV trajectory design problem in a decentralized manner.
In specific, we consider the scenario where multiple UAVs in a cellular network perform different real-time sensing tasks and transmit the sensory data the BS. 
To coordinate the UAVs, we first propose a sense-and-send protocol, and solve the successful transmission probability in the protocol by using nested Markov chain. 
We then formulate the decentralized trajectory design problem based on the reinforcement learning framework.
Under the framework, we propose an enhanced multi-UAV Q-learning algorithm to solve the decentralized trajectory design problem.

In literature, most works focused on either the sensing or the transmission part in UAV networks, instead of considering UAV sensing and transmission jointly.
\textcolor[rgb]{0,0,0}{
For example, authors in \cite{tisdale2009autonomous,maza2009multi,gu2006optimal,yang2018real} focused on the sensing part.
In \cite{tisdale2009autonomous}, the autonomous path planning problem was discussed for a team of UAVs equipped with vision-based sensing system to search for a stationary target. 
In \cite{maza2009multi}, an architecture was proposed to deal with the cooperation and control of multiple UAVs with sensing and actuation capabilities for the deployment of loads.
In \cite{gu2006optimal}, the optimal cooperative estimation problem of both the position and velocity of a ground moving target is considered by using a team of UAVs.
In \cite{yang2018real}, a mobile air quality monitoring system boarded on the UAV was designed to sense the real-time air quality and estimate the air quality index maps at given location.}

\textcolor[rgb]{0,0,0}{On the other hand, authors in \cite{zhang2018joint,bor2016efficient} focused on the transmission part in UAV networks.
In \cite{zhang2018joint}, the joint trajectory and power optimization problem was formulated to minimize the outage probability in the network, in which the UAV relayed the transmission of mobile devices.
In \cite{bor2016efficient}, UAVs were used as aerial BSs which assisted the BS in providing connectivity within the cellular network, and an optimization problem was formulated to maximize the network's revenue. }
%
%However, most existing works have not considered the transmission of the real-time sensory data to the core network, which is necessary since the sensory data need to be further processed and used by the applications in the cloud or at the user end.

In \cite{zhang2018cellular}, both the sensing and transmission are taken into consideration, and an iterative trajectory, sensing, and scheduling algorithm was proposed to schedule UAVs' trajectories in a centralized manner, in which the task completion time was minimized.
Nevertheless, the decentralized trajectory design problem remains to be lack of discussion, which is important since in practical scenarios the UAVs may belong to different entities, and thus having the incentives to maximize their own utilities.

In this paper, the main contributions can be summarized as follows.
\begin{itemize}
	\item We propose a sense-and-send protocol to coordinate UAVs performing real-time sensing tasks, and solve the probability for successful valid sensory data transmission in the protocol by using nested Markov chains.
	\item We adopt the reinforcement learning framework for the UAV trajectory design problem, based on which an enhanced multi-UAV Q-learning algorithm is proposed to solve the problem in a decentralized manner.
	\item Simulation results show that the enhanced multi-UAV Q-learning algorithm converges faster and to higher rewards of UAVs compared to both single-agent and opponent modeling Q-learning algorithms.
\end{itemize}

The rest of this paper is organized as follows.
In Section II, the system model is described. 
In Section III, we propose the sense-and-send protocol to coordinate the UAVs performing real-time sensing tasks.
We analyze the performance of the proposed sense-and-send protocol in Section IV, and solve the successful transmission probability by using nested Markov chains. 
Following that, the reinforcement learning framework and the enhanced multi-UAV Q-learning algorithm are given in Section V, together with the analyses of complexity, convergence, and scalability. 
The simulation results are presented in Section VI. 
Finally, the conclusions are drawn in Section VII.
%\vspace{-0.2cm}
%%%%%%%%%%%%%%%%%%%%%%%%%%%%%%%%%%%%%%%%%%%%%%%
\section{System Model} \label{sec: system model}
\begin{figure}[!t] % Example image
	\center{\includegraphics[width=1\linewidth] {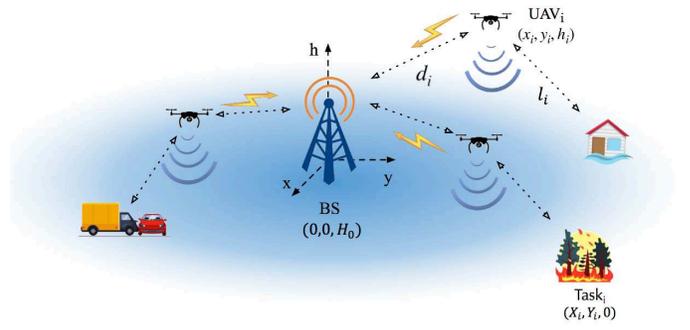}}
	\caption{Illustration on the single-cell UAV network, in which UAVs perform real-time sensing tasks.}
	\label{fig: example model}
\end{figure}

%
%%%%%%%%%%%%%%%%%%%%%%%%%%%%%%%%%%%%%%%%%%%%%%%
\textcolor[rgb]{0,0,0}{As illustrated in Fig.~\ref{fig: example model},} we consider a single cell orthogonal frequency-division multiple access (OFDMA) network which consists of $N$ UAVs to perform real-time sensing tasks. 
Setting the horizontal location of the BS to be the origin of coordinates, the BS and UAVs can be specified by 3D cartesian coordinates, i.e., the $i$-th UAV can be denoted as $s_i=(x_i,y_i,h_i)$, and the BS can be denoted as $(0,0,H_0)$ with $H_0$ being its height.
%!The following two sentences needed to be reformed to avoid plagiarism.
The UAV $i$ performs its real-time sensing task $i$, the location of which is denoted as $(X_i,Y_i,0)$. 
\textcolor[rgb]{0,0,0}{To perform real-time sensing task, each UAV continuously senses the condition of its task, and sends the collected sensory data to the BS immediately.
In this regard, the sensing process and the transmission process jointly determine the UAVs' performance on the real-time sensing tasks.}
The sensing and transmission models for the UAV are described in the following. 

%In this paper, we jointly consider the sensing process and the uplink transmission process and propose a distributed trajectory control algorithm of the UAV as well as an uplink SC allocation algorithm of the BS in order to minimize the delay for the task to be known successfully by the BS.
%The model of UAV sensing, transmitting, the actions in performing sensing tasks are described respectively in the following subsections. 
%Following that the mechanism of BS uplink SC allocation is exhibited.

\subsection{UAV Sensing}
To evaluate the sensing quality of the UAV, we utilize the probabilistic sensing model as introduced in \cite{Shakhov2017Experiment,Chakraborty2013Network}, where the successful sensing probability is an exponential function of the distance between the UAV and its task.
Supposing that UAV $i$ senses task $i$ for a second, the probability for it to sense the condition of its task successfully can be expressed as
\beq \label{equ: successful task-sensing probability}
Pr_{s,i}= e^{-\lambda l_{i}},
\eeq
in which $\lambda$ is the parameter evaluating the sensing performance and $l_i$ denotes the distance between UAV $i$ and its sensing task $i$. 

It is worth noticing that UAV $i$ cannot figure out whether the sensing is successful or not from its collected sensory data, due to its limited on-board data processing ability. 
Therefore, UAV $i$ needs to send the sensory data to the BS, and leaves for the BS to decide whether the sensory data is valid or not.
%If the sensing is successful, UAV $i$ will collect valid sensory data of task $j$.
Nevertheless, UAV $i$ can evaluate its sensing performance by calculating the successful sensing probability based on (\ref{equ: successful task-sensing probability}).

\subsection{UAV Transmission}\label{sec: UAV transmission}
In the UAV transmission, the UAVs transmit the sensory data to the BS over orthogonal subchannels (SCs) to avoid interference.
We adopt the 3GPP channel model for evaluating the urban macro cellular support for UAVs \cite{3GPP2017R14,3GPP2017R15}.

Denoting the transmit power of UAVs as $P_u$, the received signal-to-noise ratio (SNR) at the BS of UAV $i$ can be expressed as
\beq \label{Equ: received power of BS}
\gamma_{i} = \frac{P_u\|H_i\|}{N_010^{\mathrm{PL}_{a,i}/10}},
\eeq
in which $\mathrm{PL}_{a,i}$ denotes the air-to-ground pathloss,
$N_0$ denotes the power of noise at the receiver of the BS,
and $H_i$ is the small-scale fading coefficient.
Specifically, the pathloss $\mathrm{PL}_{a,i}$ and small-scale fading $H_i$ should be calculated in two cases separately, i.e., line-of-sight (LoS) case and non LoS (NLoS) case.
The probability for the channel UAV $i$-BS to contain a LoS component is denoted as $Pr_{Los,i}$, and can be calculated as
\beq
Pr_{\mathrm{LoS},i}=\begin{cases}
	1,& r_i\leq r_c,\\
	\frac{r_c}{r_i} + e^{-r_i/p_0+r_c/p_0}, & r_i>r_c,
\end{cases},
\eeq
in which $r_i = \sqrt{x_i^2+y_i^2}$, $p_0 = 233.98\log(h_i) - 0.95$, and $r_c =\max\{ 294.05\log_{10}(h_i) - 432.94, 18\}$.

When the channel contains a LoS component, the pathloss from UAV $i$ to the BS can be calculated as $\mathrm{PL}_{a,i}=\mathrm{PL}_{\mathrm{LoS},i}=30.9+(22.25-0.5\log_{10}(h_i))\log(d_i)+20\log_{10}(f_c)$, where $f_c$ is the carrier frequency and $d_i$ is the distance between the BS and UAV $i$.
In the LoS case, the small-scale fading $H_i$ obeys Rice distribution with scale parameter $\Omega = 1$ and shape parameter $K [\mathrm{dB}]= 4.217\log_{10}(h_i)+5.787$.
On the other hand, when the channel contains none LoS components, the pathloss from UAV $i$ to the BS can be calculated as $\mathrm{PL}_{a,i}=\mathrm{PL}_{\mathrm{NLoS},}= 32.4+(43.2-7.6\log_{10}(h_i))\times\log_{10}(d_i)+20\log_{10}(f_c)$, and the small-scale fading $H_i$ obeys Rayleigh distribution with zero means and unit variance.

%The noise at the BS satisfies the Gaussian distribution with zero mean and $N_0$ as variance.
%Based on the above notation, the received signal of BS can be expressed by
%\beq
%S_R(t) = \sqrt{P_{BS,i}(t)}  S_T(t) + n(t),
%\eeq
%where $S_T(t)$ is the signal of unit energy from the UAV, and $n(t)$ 
%is the noise received at the BS, which satisfies the Gaussian distribution with zero mean and $N_0$ as variance. 
To achieve a successful transmission, the SNR at the BS needs to be higher than the decoding threshold $\gamma_{th}$, otherwise, the uplink transmission is failed. 
Therefore, each UAV can evaluate its probability of successful uplink transmission by calculating the probability for the SNR at BS to be larger than $\gamma_{th}$.
The successful uplink transmission probability $Pr_{\mathrm{Tx},i}$ for UAV $i$ can be calculated as 
\begin{align} \label{equ: successful tx probability}
&Pr_{\mathrm{Tx},i} =\nonumber \\
&~~~Pr_{\mathrm{Los},i}(1\!-\!F_{ri}(\chi_{\mathrm{LoS},i}) )\!+\!(1\!-\!Pr_{\mathrm{LoS},i})(1\!-\!F_{ra}(\chi_{\mathrm{NLoS},i})),
\end{align}
in which $\chi_{\mathrm{NLoS},i}= N_0 10^{0.1\mathrm{PL}_{\mathrm{NLoS},i}}\gamma_{th}/P_u, ~\chi_{\mathrm{LoS},i}= N_0 10^{0.1\mathrm{PL}_{\mathrm{LoS},i}}\gamma_{th}/P_u$, $F_{ri}(x) =1- Q_1(\sqrt{2K},x\sqrt{2(K+1)})$ is the cumulative distribution function (CDF) of the Rice distribution with $\Omega = 1$\cite{rice1944mathematical}, and $F_{ra}(x) = 1-e^{-x^2/2}$ is the CDF of the Rayleigh distribution with unit variance. Here $Q_1(x)$ denotes the Marcum Q-function of order 1 \cite{marcum1950table}.

\section{Sense-and-Send Protocol}
In this section, we propose a sense-and-send protocol to coordinate the UAVs performing the sensing tasks.
We first introduce the sense-and-send cycle, which consists of the beaconing phase, the sensing phase and the transmission phase. 
After that, we describe the uplink SC allocation mechanism of the BS.
% and describe the uplink SC allocation mechanism in the protocol.
%After that, we calculate the probability of valid sensory data transmission by using a nested bi-level Markov chain. 
%Finally, the flying direction selection is specified and the distributed trajectory design of the UAVs is formulated by using a reinforcement learning framework.  
%describe the uplink SC allocation mechanism adopted by the BS.
%After that, we calculate the expectation delay of valid sensory data transmission by using a nested bi-level Markov chain. Finally, the distributed trajectory design of the UAVs are formulated as a dynamic game.  

\subsection{Sense-and-Send Cycle} \label{sec: sense-and-send cycle}
\begin{figure}[!t] % Example image
	\center{\includegraphics[width=0.8\linewidth] {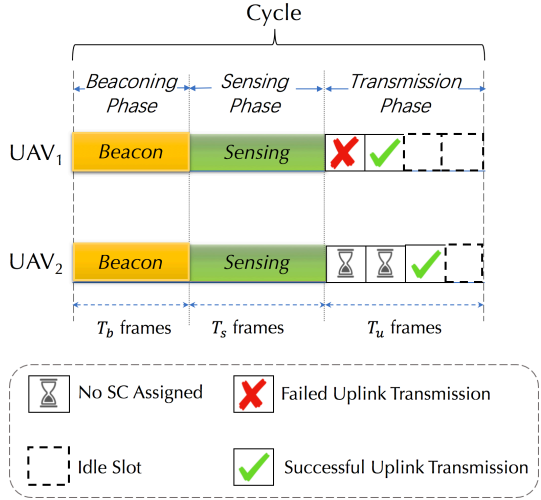}}
	\caption{Illustration on the sense-and-send protocol.}
	\label{fig: protocol}
\end{figure}
In this paper, we propose that the UAVs perform the sensing tasks in a synchronized iterative manner.
Specifically, the sensing process is divided into cycles indexed by $k$.
In each cycle, each UAV senses its task and then reports the collected sensory data to the BS for data processing.
\textcolor[rgb]{0,0,0}{
In order to synchronize the transmissions of the UAVs, we further divide each cycle into \emph{frames}, which serves as the basic time unit for SC allocation.
In specific, we assume that the collected sensory data of each UAV in a cycle can be converted into a single data frame with the same length, and the duration of the transmission and acknowledgement of that data frame is denoted as a frame.
%Therefore, in each cycle, the basic time unit for the BS to allocate the uplink sub-channel resources for the UAV sensory data transmission is $n$ time slots, which is denoted as a \emph{frame}.
Based on that, we denote the number of frames that contained in each cycle as  $T_c$. 
}

\textcolor[rgb]{0,0,0}{The cycle consists of three separated phases, i.e., the \emph{beaconing phase}, \emph{sensing phase} and the \emph{transmission phase}, which contain $T_b$, $T_s$ and $T_u$ frames, respectively.
The duration of the beaconing phase and sensing phase is considered to be fixed and determined by the time necessary in transmitting beacon frames and collecting sensory data. 
On the other hand, the duration of the transmission phase is decided by the BS considering the network conditions.}
% as in \cite{Zhang2018Cooperative}.
%Specifically, we propose the UAVs to perform the sensing tasks by following the \emph{sense-and-send protocol}.
\textcolor[rgb]{0,0,0}{As illustrated in Fig.~\ref{fig: protocol}, we consider that the sensing and transmission phases are separated to avoid the possible interference between them.}\footnote{
\textcolor[rgb]{0,0,0}{For example, the UAV's transmission will interfere with its sensing if the UAV tries to sense the electromagnetic signal in the nearby frequency band of its transmission. }}
%Moreover, since the power provided by the UAV's battery is limited, it may not be able to support simultaneous sensing and transmission.}

\textcolor[rgb]{0,0,0}{
In the beaconing phase, each UAV sends the its location to the BS on its beacon through the control channel, which can be obtained by the UAV from the GPS positioning.
Collecting the beacon frames sent by the UAVs, the BS then broadcasts to inform the UAVs of the general network settings as well as the locations of all the UAVs.
By this means, UAVs obtain the locations of other UAVs in the beginning of each cycle.
Based on the acquired information, each UAV then decides its flying trajectory in the cycle and informs the BS by transmitting another beacon.}

In the sensing phase, each UAV senses the task for $T_s$ frames continuously, during which it collects the sensory data. 
In each frame of the transmission phase, the UAVs attempt to transmit the collected sensory data to the BS.
In specific, there are four possible situations for each UAV which are described as follows.
\begin{itemize}
	\item \textbf{No SC Assigned}: In this case, UAV $i$ is not assigned any uplink SCs by the BS, and therefore cannot transmit its collected sensory data to the BS.  It will wait for the BS to assign a SC to it to transmit sensory data.
	\item \textbf{Failed Uplink Transmission}: In this case, UAV $i$ is assigned an uplink SC by the BS, however the transmission is unsuccessful due to the low SNR at the BS. Therefore, UAV $i$ attempts to send the sensory data again to the BS in the next frame. 
	\item \textbf{Successful Uplink Transmission}: In this case, UAV $i$ is assigned an uplink SC by the BS, and it succeeds in sending its collected sensory data to the BS.
	\item \textbf{Idle Frame}: In this case, UAV $i$ has successfully sent its sensory data in the former frames, and will keep idle in the rest of the cycle until the beginning of the next cycle.
\end{itemize}

Note that in the model we have assumed that the transmission of sensory data occupies a single frame. 
Nevertheless, it can be extended to the case where the sensory data transmission takes $n$ frames straightforwardly.
In that case, the channel scheduling unit becomes $n$ frames instead of a single frame.

\subsection{Uplink Subchannel Allocation Mechanism} \label{sec: uplink sc allocation mechanism}
Since the uplink SC resources are usually scarce, thus in each frame of the transmission phase, there may exist more UAVs requesting to transmit their sensory data than the number of available uplink SCs.
To deal with this problem, the BS adopts the following SC allocation mechanism to allocate the uplink SCs to the UAVs.

%In this paper, we assume that the BS allocates the uplink SCs according to the successful transmission probabilities of the UAVs, which are the probabilities for UAVs to transmit the sensory data to the BS successfully in the frame.
%The successful transmission probability of UAV $i$ in frame $t$ of $k$-th cycle can be denoted as $Pr_{Tx,i}^k(t)$, which can be calculated based on (\ref{equ: successful tx probability}).

In each frame, the BS allocates the $C$ available uplink SCs to the UAVs with uplink requirements, in order to maximize the sum of successful transmission probabilities of uplink UAVs. 
Based on the matching algorithm in \cite{Demange1986Multi}, it is equivalent that the BS allocates the $C$ available SCs to the first $C$ UAVs with the highest successful transmission probabilities in the frame.
\textcolor[rgb]{0,0,0}{The successful transmission probabilities of UAVs can be calculated by the BS based on (\ref{equ: successful tx probability}), using the information on the trajectories of the UAVs collected in the beaconing phase.}
Moreover, denoting the transmission state of the UAVs in the $k$-th cycle as the vector $\bm I^{(k)} (t) $, \textcolor[rgb]{0,0,0}{$\bm I^{(k)} (t) = (I^{(k)}_1(t),...,I^{(k)}_ N(t))$. 
Here, $I_i^{(k)}(t) = 0$ if UAV $i$ does not succeed in transmitting its sensory data to the BS at the beginning of the $t$-th frame, otherwise, $I_i^{(k)}(t)=1$.}
Based on the above notations, the uplink SC allocation can be expressed by the channel allocation vector $\bm\nu^{(k)}(t) = (\nu_1^{(k)}(t),...,\nu_N^{(k)}(t))$, in which the elements can be expressed as follows.
\begin{equation} \label{equ: channel allocation indicator}
\nu_{i}^{(k)}(t) = \begin{cases}
	1, &  Pr_{\mathrm{Tx},i}^{(k)}(t) I_{i}^{(k)}(t)\geq (\bm{Pr}_{\mathrm{Tx}}^{k}(t)\bm I^{(k)}(t))_C,\\
	0,& o.w.
\end{cases}.
\end{equation}

Here, $\nu_i^{(k)}(t)$ is the \emph{channel allocation indicator} for UAV $i$, i.e., $\nu_i^{(k)}(t)=1$ only if an uplink SC is allocated to UAV $i$ in the $t$-th frame, 
$Pr_{\mathrm{Tx},i}^{(k)}(t)$ denotes the successful transmission probability of UAV $i$ in frame $t$ of $k$-th cycle,
and $(\bm{Pr}_{\mathrm{Tx}}^{(k)}(t)\bm I^{(k)}(t))_C$ denotes the $C$-th largest successful transmission probabilities among the UAVs who have not succeeded in uploading sensory data before the $t$-th frame.
%where $Pr_{ss,i}(t)$ denotes the probability for UAV $i$ to have valid sensory data in frame $t$. 
%To describe $Pr_{ss,i}(t)$ clearly, we divide the time line into segments according to each sensing cycle, and denote the $t$-th frame in each sensing cycle as frame $t$.
%Therefore, the sensing process consists of the first $T_s$ frames and the uplink transmitting process consists of the next $T_u$ frames. 
%Moreover, since the change of location of UAVs during each cycle is small, therefore, we assume the location of each UAV is fixed within each cycle, and index the cycle by $k$. 
%Therefore, the valid sensory data transmission probability can be calculated as follows
%\beq
%Pr_{sTx,i}(t) = \big(1 - (1-e^{-\lambda l_i(k)})^{T_s}\big)\times .
%\eeq

Since the location of UAV $i$ determines UAV $i$'s distance to the BS, it influences the successful uplink transmission probability.
As the UAVs which have larger successful transmission probabilities are more likely to be allocated SCs, the UAVs have the motivation to compete with each other by selecting trajectories where they have higher probabilities to be allocated SCs. 
Consequently, the UAVs need to design their trajectories with the consideration of their distance to the BS and the task, as well as the trajectories of other UAVs.
%To tackle with this issue, we adopt the reinforcement learning to solve the trajectory design problems for the UAVs which will be analyzed in detail in Section \ref{sec: reinforcement learning in UAV trajectory design}.

\section{Sense-and-Send Protocol Analysis} \label{sec: analysis on reward}
%\subsection{Probability for valid sensory data Transmission}
In this section, we analyze the performance of the proposed sense-and-send protocol by calculating the probability of successful valid sensory data transmission, which plays an important role in solving the UAV trajectory design problem.
We first specify the state transition of UAVs in the sensing task by using nested bi-level Markov chains.
The outer Markov chain depicts the state transition of UAV sensing, and the inner Markov chain depicts the state transition of UAV transmission, which will be described in the following parts, respectively.

\subsection{Outer Markov Chain of UAV Sensing}
\begin{figure}[!t] % Example image
	\center{\includegraphics[width=0.9\linewidth] {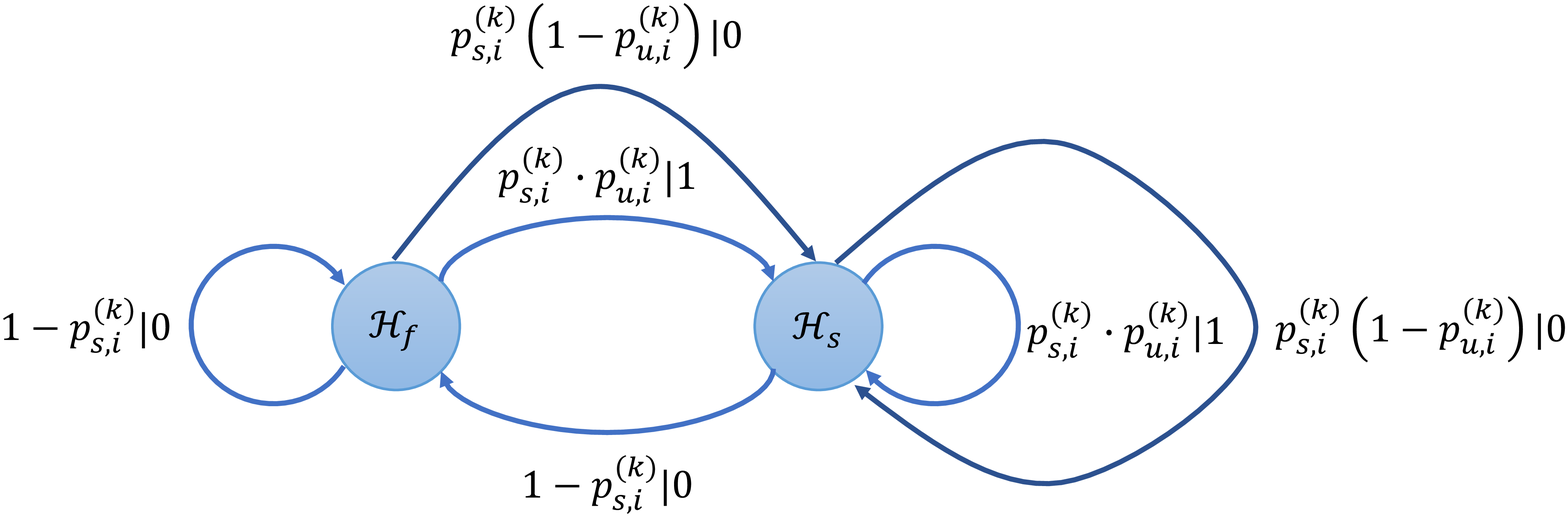}}
	\caption{Illustration on outer Markov chain of UAV sensing.}
	\label{fig: outer markov chain}
\end{figure}

In the outer Markov chain, the state transition takes place among different cycles.
As shown in Fig.~\ref{fig: outer markov chain}, for each UAV, it has two states in each cycle, i.e., state $\mathcal H_f$ to denote that the sensing is failed, and state $\mathcal H_s$ to denote that the sensing is successful.
Supposing the successful sensing probability of UAV $i$ in the $k$-th cycle is $p_{s,i}^{(k)}$, UAV $i$ transits to the $\mathcal H_s$ state with probability $p_{s,i}^k$ and transits to the $\mathcal H_f$ state with probability $(1-p_{s,i}^{(k)})$ after the $k$-th cycle. 
The value at the right side of the transition probability denotes the number of  valid sensory data that have been transmitted successfully to the BS in the cycle.

Besides, we denote the probability for UAV $i$ to successfully transmit the sensory data to the BS as $p_{u,i}^{(k)}$. 
Therefore, UAV $i$ successfully transmits valid sensory data to the BS with the probability $p_{s,i}^{(k)}  p_{u,i}^{(k)}$, and with probability $p_{s,i}^{(k)}(1-p_{u,i}^{(k)})$, no valid sensory data is transmitted to the BS though the sensing is successful in the $k$-th cycle.
The probability $p_{u,i}^{(k)}$ can be analyzed by the inner Markov chain of UAV transmission in the next subsection, and $p_{s,i}^{(k)}$ can be calculated as follows.

%Supposing that at the beginning of the $k$-th cycle, UAV $i$ is at location $(x_i^k(1),y_i^k(1), h_i^k(1))$, and it will move to the next location $(x_i^{k+1}(1),y_i^{k+1}(1),h_i^{k+1}(1))$.
Since the change of UAVs' locations during each frame is small, we assume that the location of each UAV is fixed within each frame. 
Therefore, the location of UAV $i$ in the $k$-th cycle can be expressed as a function of the frame index $t$, i.e., $s_i^{(k)}(t)=(x_i^{(k)}(t),y_i^{(k)}(t),h_i^{(k)}(t)),~t\in[1,T_c]$.
Similarly, the distance between UAV $i$ and its task can be expressed as $l_{i}^{(k)}(t)$, and the distance between the UAV and the BS can be expressed as $d_i^{(k)}(t)$.
Moreover, we assume that the UAVs move with uniform speed and fixed direction in each cycle after the beginning of the sensing phase.
Therefore, at the $t$-th frame of the $k$-th cycle, the location of UAV $i$ is 
\textcolor[rgb]{0,0,0}{
\begin{equation}
s_i^{(k)}(t) = s_i^{(k)}(T_b) +\frac{t}{T_c}(s_i^{(k+1)}(1)-s_i^{(k)}(T_b) ), ~t\in[T_b,T_c]. 
\end{equation}}
Since each UAV senses for the first $T_s$ frames in the $k$-th cycle, the successful sensing probability of UAV $i$ in the cycle can be calculated as
\beq
p_{s,i}^{(k)} = \prod_{t=T_b+1}^{T_s+T_b} (Pr_{s,i}^{(k)}(t))^{t_f} = \prod_{t=T_b+1}^{T_s+T_b} e^{-\lambda t_f l_i^{(k)}(t)}.
\eeq
in which $t_f$ denotes the duration of a frame.

\subsection{Inner Markov Chain of UAV Transmission}
\begin{figure}[!t] % Example image
	\center{\includegraphics[width=1\linewidth] {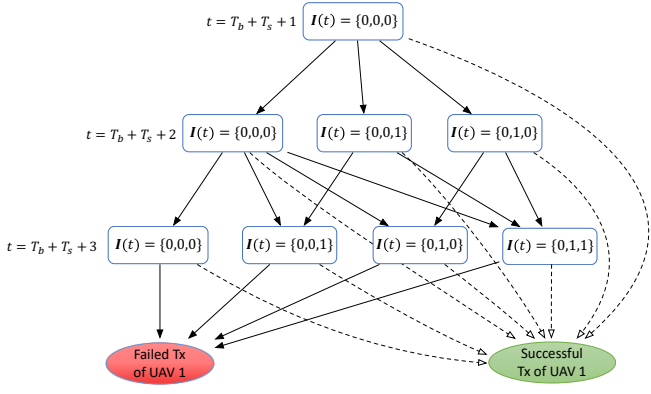}}
	\caption{Illustration on inner Markov chain of UAV 1's transmission given $C=1,~ N=3, ~T_u=3$.}
	\label{fig: inner markov chain}
\end{figure}
For simplicity, we omit the superscript $k$ indicating the index of the cycle. 
Since the general state transition diagram is rather complicated, we illustrate the inner Markov chain by giving an example where the number of available uplink SC $C=1$, the number of UAVs $N=3$ and the number of uplink transmission frames $T_u = 3$.

Taking UAV 1 as an example, the state transition diagram is given in Fig.~\ref{fig: inner markov chain}. 
The state of the UAVs in frame $t$ can be represented as the transmission state vector $\bm I(t)$ as defined in Section \ref{sec: uplink sc allocation mechanism}.
Initially $t=T_b+T_s+1$, the transmission state is $\bm I(T_b+T_s+1) =\{0,0,0\} $, which indicates that UAVs 1, 2, and 3 have not succeeded in uplink transmission at the beginning of the transmission phase, and all of them are competing for the uplink SCs.
In the next frame, the transmission state will transit to the \emph{Successful Tx} state for UAV 1, if the sensory data of UAV 1 has been successfully transmitted to the BS.
The probability for this transition equals to $Pr_{Tx,1}(T_b+T_s+1) \nu_1(T_b+T_s+1)$, i.e., the probability for successful uplink transmission if a SC is allocated to UAV 1, otherwise, it equals to zero.

However, if UAV 1 does not succeed in uplink transmission, the transmission transits into other states, which is decided by whether other UAVs succeed in uplink transmission, e.g., it transits to $\bm I(T_b+T_s+2) = (0,0,1)$ if UAV 3 succeeds in the first transmission frame. 
Note that when other UAVs succeed in transmitting sensory data in the previous frames, UAV 1 will face less competitors in the following frames, and thus, it have a larger probability to transmit successfully.
Finally, when $t=T_c$, i.e., the last transmission frame in the cycle, UAV 1 will enter the \emph{Failed Tx} state if it does not transmit the sensory data successfully, which means that the sensory data in this cycle is failed to be uploaded.
Therefore, to obtain the $p_{u,i}$ in the outer Markov chain, it is equivalent to calculate the absorbing probability of successful Tx state in the inner Markov chain.

From the above example, it can be observed that the following general recursive equation holds for UAV $i$ when $t\in[T_b+T_s+1,T_c]$,
\begin{align} \label{equ: recursive equation in suc. tx. prob.}
&Pr_{u,i}\{t|\bm I(t)\} =  Pr_{\mathrm{Tx},i}(t)\nu_i(t)  \nonumber\\&~~~
+\!\sum_{\substack{\bm I(t+1),\\I_i(t+1) = 0}}Pr\{\bm I(t+1)|\bm I(t)\}Pr_{u,i}\{t+1|\bm I(t+1)\},
\end{align}
in which $Pr\{\bm I(t+1)|\bm I(t)\}$ denotes the probability for the transmission state vector of the $(t+1)$-th frame to be $\bm I(t+1)$ given that of the $t$-th frame to be $\bm I(t)$, 
and $Pr_{u,i}\{t|\bm I(t)\}$ denotes the probability for UAV $i$ to transmit sensory data successfully after the $t$-th frame in the current cycle, given the transmission state $\bm I (t) (I_i(t)=0)$.

Since the successful uplink transmission probabilities of the UAVs are independent, we have $Pr\{\bm I(t+1)| \bm I(t) \} = 	\prod_{i=1}^N Pr\{I_i(t+1)|I_i(t)\}$, in which $Pr\{I_i(t+1)|I_i(t)\}$ can be calculated as follows.
\beq
\begin{cases}
&Pr\{I_i(t+1) = 0|I_i(t) = 0\} = 1-Pr_{\mathrm{Tx},i}(t),\\
&Pr\{I_i(t+1) = 1|I_i(t) = 0\} = Pr_{\mathrm{Tx},i}(t),\\
&Pr\{I_t(t+1) = 0|I_i(t) = 1\} = 0,\\
&Pr\{I_t(t+1) = 1|I_i(t) = 1\} = 1.
\end{cases}
\eeq
Here, the first two equations hold due to that the successful transmission probability in the $t$-th frame is $Pr_{\mathrm{Tx},i}(t)$.
The third and forth equations indicate that the UAVs keep idle in the rest of frames once they have successfully sent their sensory data to the BS.

Based on equation (\ref{equ: recursive equation in suc. tx. prob.}), the recursive algorithm can be used to solve $Pr_{u,i}\{t|\bm I(t)\}$, which is described in Alg. \ref{alg: solve successful transmission probability}.
Therefore, the successful transmission probability can be obtained by $p_{u,i} = Pr_{u,i}\{T_b+T_s+1|\bm I(T_b+T_s+1)\}$.
In summary, the successful valid sensory data transmission probability for UAV $i$ in the $k$-th cycle can be calculated as
\begin{equation} \label{equ: successful valid sensory data transmission prob.}
p_{\mathrm{sTx},i}^{(k)} =p_{s,i}^{(k)} p_{u,i}^{(k)}.	
\end{equation}

\subsection{\textcolor[rgb]{0,0,0}{Analysis on Spectrum Efficiency}}
In this paper, we evaluate the spectrum efficiency by the average number of valid sensory data transmissions per second, which is denoted as $N_{vd}$. 
The value of $N_{vd}$ is influenced  by many aspects, such as the distance between the BS and the tasks, the number of available SCs, the number of UAVs in the network, and the duration of the transmission phase.

In this paper, we analyze the influence of the duration of transmission phase $T_u$ on $N_{vd}$ in a simplified case.
%In this way, we are able to derive some qualitative results and avoid the complicated analysis on the successful transmission probability.
Assuming all the UAVs are equivalent, i.e., they have the same probabilities for successful uplink transmission in a frame, the same probabilities for successful sensing and the same probability to be assigned sub-channels.
Based on the above assumptions, the following proposition can be derived.
\begin{proposition}\label{prop: theo res}
\emph{(Optimal duration of transmission phase)}
	When the UAVs are equivalent, and have the probability for successful sensing $p_s$, the probability for successful uplink transmission $p_u$, then $N_{vd}$ first increases then decreases with the increment of $T_u$, and the optimal $T_u^*$ can be calculated as 
	\begin{equation}
		T_u^* =\frac{N}{C\ln (1-p_u)}(1+W_{-1}(-\frac{(1-p_u)^{\frac{CT_u}{N}}}{e}))-T_b-T_s,
	\end{equation}
	in which $W_{-1}(\cdot)$ denotes the lower branch of Lambert-W function \cite{Corless1996}.
\end{proposition}

\indent {\emph{Proof:}} See Appendix \ref{appx: a}.$\hfill\blacksquare$

The above proposition in special case sheds light on the relation between spectrum efficiency and duration of transmission phase in general cases. 
In general cases where the UAVs are not equivalent, the spectrum efficiency also first increases then decreases with the duration of transmission phase. 
This is because when $T_u  = 0$, $N_{vd} = 0$, and when $T_u \rightarrow \infty$, $N_{vd}\rightarrow 0$.
	
\begin{algorithm}[t]  
  \caption{Algorithm for successful transmission probability in a cycle.}  
  \label{alg: solve successful transmission probability}
  \begin{algorithmic}[1]  
    \REQUIRE  
    Frame index ($t$); Channel state vector at current frame ($\bm I(t)$);Length of beaconing phase ($T_b$); Length of sensing phase ($T_s$); Length of transmission phase ($T_u$); Location of UAVs in each frame ($(x_i(t'),y_i(t'),h_i(t')),~\forall t'=T_b+T_s+1,...,T_c, i=1,...,N$); Number of channels ($C$). 
    \ENSURE  
      $Pr_{u,i}\{t|\bm I(t)\},~i=1,...,N$;  
    \IF{$t=T_s+1$}
    \STATE Set $\bm I(t) := \bm 0$, $Pr_{u,i}\{t|\bm I(t)\} := 0, i=1,...,N$;
	\ELSIF{$t >T_b+T_s+T_u$}
	\RETURN $Pr_{u,i}\{t|\bm I(t)\} = 0,~i=1,...,N$.
	\ENDIF
    \STATE Calculate the successful uplink transmission probabilities $Pr_{\mathrm{Tx},i}(t)$ of each UAV $i$ in current frame $t$ based on (\ref{equ: successful tx probability}).
    \STATE Determine the SC allocation indicator $\bm\nu(t)$ based on (\ref{equ: channel allocation indicator}).
    \FOR{$I_i(t) = 0$}
    \STATE $Pr_{u,i}\{t|\bm I(t)\} := Pr_{Tx,i}(t)\nu_i(t)$.
    \ENDFOR
    \FOR{all $\bm I(t+1)$ with $Pr\{\bm I(t+1)|\bm I(t)\}>0$}
	\STATE Solve $Pr_{u,i}\{t+1|\bm I(t+1)\}$ by calling Alg. \ref{alg: solve successful transmission probability}, in which $t := t+1$ and $\bm I(t):=\bm I(t+1)$ and other parameters hold.
	\STATE
	$Pr_{u,i}\{t|\bm I(t)\} \!:=\!  Pr_{u,i}\{t|\bm I(t)\}  \newline ~ ~~~~~~~~~~+Pr\{\bm I(t+1)|\bm I(t)\} Pr_{u,i}\{t+1|\bm I(t+1)\}.$
	\ENDFOR
	\RETURN $Pr_{t,i}\{t|\bm I(t)\},~i=1,...,N$.
	\end{algorithmic}  
\end{algorithm}

\section{Decentralized Trajectory Design} \label{sec: reinforcement learning in UAV trajectory design}
In this section, we first describe the decentralized trajectory design problem of UAVs, and then formulate a reinforcement learning framework for the UAVs to determine their trajectories. 
After that, we describe the single-agent and multi-agent reinforcement learning algorithms under the framework, and proposed an enhanced multi-UAV Q-learning algorithm to solve the UAV trajectory design problem.  

\subsection{UAV Trajectory Design Problem}
\begin{figure}[!t] % Example image
	\center{\includegraphics[width=0.45\linewidth] {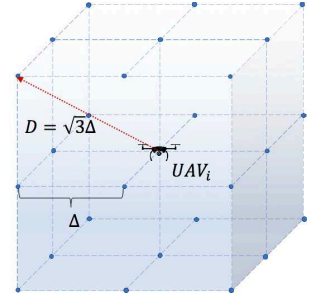}}
	\caption{Illustration on the set of available spatial points that the UAV $i$ can reach in the next cycle.}
	\label{fig: spatial cube}
\end{figure}

Before the formulation of the trajectory design problem, we first set up a model for the UAV trajectory.
In this paper, we focus on the cylindrical region with the maximum height $h_{\max}$ and the radius of the cross section $R_{\max}$ which satisfies $R_{\max} = \max \{R_i \big | R_i = \sqrt{X_i^2+Y_i^2},\forall i\in[1,N] \}$, since it is inefficient for the UAVs to move further than the farthest task. 
Moreover, we assume that the space is divided into a finite set of discrete spatial points $\mathcal S_p$, which is arranged in a square lattice pattern as shown in Fig.~\ref{fig: spatial cube}.

To obtain the trajectories of the UAVs, we assume that the UAVs can select their flying directions at the beginning of each cycle. 
For example, UAV $i$ locates at the spatial point $s_i^{(k)} = (x_i^{(k)}, y_i^{(k)}, h_i^{(k)})\in \mathcal S_p$ at the beginning of the $k$-th cycle, and decides which spatial point it will move to next, which is equivalent to determining its flying direction in this cycle.
After the UAV has selected its flying direction, it will move along the direction towards the destination point with a uniform speed in this cycle. 
%
%Moreover, since the change of UAVs' locations during each frame is small, we assume that the location of each UAV is fixed within a frame. 
%Therefore, the location of UAV $i$ in the $k$-th cycle can be expressed as a function of the frame index $t$, i.e., $(x_i^k(t),y_i^k(t),h_i^k(t))$.
%Similarly, the distance between UAV $i$ and its task can be expressed as $l_{i}^k(t)$, and the distance between the UAV and the BS can be expressed as $d_i^k(t)$.

The available spatial points that UAV $i$ can reach is within the maximum distance it can fly in a cycle, which is denoted as $D$.
Assuming that the distance between two adjacent spatial points is $\Delta=D/\sqrt 3$, and thus, the available spatial point UAV $i$ can fly to in the $k+1$ cycle is within a cube centered at $(x_i^{(k)}, y_i^{(k)}, h_i^{(k)})$ with the length of side $2\Delta$, as illustrated in Fig~\ref{fig: spatial cube}.
It can be seen that there are at most $27$ available flying directions can be selected by the UAVs in each cycle. 
We denote the set of all the vectors from the center to the available spatial points in the cube as the available action set $\mathcal A$ of the UAVs.
However, it is worth noticing that when the UAV is at the marginal location (e.g., flying at the minimum height), there are less available actions to be selected. 
To handle the differences among the available action sets at different spatial points, we denote the available action set at the spatial point $s$ as $\mathcal A(s)$.

In this paper, we consider the reward of each UAV to be the number of successful valid sensory data transmissions in the previous cycles.
Therefore, the UAVs have incentive to maximize the total amount of successful valid sensory data transmission by designing their trajectories. 
Besides, we assume that the UAVs have discounting valuation on the successfully transmitted valid sensory data.
For the UAVs in the $k$-th cycle, the successfully valid sensory data transmitted in the $k'$-th cycle is worth only $\rho^{|k'-k|}~(\rho\in[0,1))$ the successful valid sensory data transmitted in the current cycle, due to the timeliness requirements of real-time sensing tasks. 
%If $\rho = 0$, the UAV is short-sighted, that is, it only maximize the immediate rewards. 
%On the other hand, as $\rho$ approaches 1, the UAV becomes more far-sighted.
Therefore, at the beginning of $k$-th cycle, the expected sum of discounted rewards of UAV $i$ can be denoted as $
	G_i^{(k)} = \sum_{n=0}^\infty \rho^{n}R_{i}^{(k+n)}$, where $R_i^{(k)} = 1$ if valid sensory data is successfully transmitted to the BS by UAV $i$ in the $k$-th cycle, otherwise, $R_i^{(k)} = 0$.
Based on the above assumptions, the UAV trajectory design problem can be formulated as 
\begin{align}\label{opt: trajectory design problem}
\max_{a_i^{(k)}\in \mathcal A}\quad & \sum_{n=0}^\infty \rho^{n}R_{i}^{(k+n)},\\
{s.t.} \quad & s_i^{(k)}+a_i^{(k)}\in \mathcal S_p, \tag{\ref{opt: trajectory design problem}a}
\end{align}

%Here we omit the possible location change within each cycle for that the duration of each cycle is small.
%Therefore, we assume that the location of each UAV in each cycle is same, i.e., the UAV changes its location at the beginning of each cycle.
%In the following part of the section, we formulate the distributed trajectory design problem in the UAV sensing and transmission as a stochastic game, and solve it using multi-agent reinforcement learning in the following section.

\subsection{Reinforcement Learning Framework}
Generally, the UAV trajectory design problem (\ref{opt: trajectory design problem}) is hard to solve since the rewards of the UAVs in the future cycles are influenced by the trajectories of all UAVs, which are determined in a decentralized manner and hard to model.
Fortunately, the reinforcement learning is able to deal with the problem of agent programming in environment with deficient understanding, which removes the burden of developing accurate models and solving the optimization with respect to those models.

For this reason, we adopt the reinforcement learning to solve the UAV trajectory design problem in this paper. To begin with, we formulate a reinforcement learning framework for the problem.
% for the UAVs to determine their trajectories in order to maximize their sum of valid sensory data transmission probability in a given time $T$.
%The framework of stochastic games is widely adopted to model multi-agent systems with finite states and actions \cite{Hu2003Nash}.
With the help of \cite{Yang2004Multiagent}, the reinforcement learning framework can be given as follows, in which the superscript $k$ is omitted for simplicity.
\begin{definition}
	A reinforcement learning framework for UAV trajectory design problem is described by a tuple $<{\mathcal{S}}_1,...,{\mathcal{S}}_{N},{\mathcal{A}}_1,...,{\mathcal{A}}_{N},\mathcal T,p_{R,1},...,p_{R,N}, \rho>$, where
	\begin{itemize}
		\item ${\mathcal{S}}_1,...,{\mathcal{S}}_N$ are finite state spaces of all the possible locations of the $N$ UAVs, and the state space of UAV $i$ equals to the finite spatial space, i.e., $\mathcal S_i = \mathcal S_p,~\forall i\in[1,N]$.
		\item ${\mathcal{A}}_1,...,{\mathcal{A}}_{N}$ are the corresponding finite sets of actions available to each agents. The set ${\mathcal{A}_i}$ consists of all the available action of UAV $i$, i.e., $\mathcal A_i = \mathcal A,~\forall i \in [1,N]$.
		\item $\mathcal T:\prod_{i=1}^N{\mathcal{S}}_i\times\prod_{i=1}^N {\mathcal{A}}_i\rightarrow ({\mathcal{S}}_p)^N$ is the state transition function. It equals to the locations of the UAVs in the next cycle for the given location profile and action profile of the UAVs in the current cycle.
		\item $p_{R,1},...,p_{R,N}: \prod_{i=1}^N{\mathcal{S}}_i\times\prod_{i=1}^N {\mathcal{A}}_i\rightarrow \Pi(0,1)^N,~i=1,...,N$ represents a reward function for each UAV. In specific, it maps the UAVs' location profile and action profile of the current cycle to the probability for UAV $i$ ($i,=1,...,N$) to get unit reward from performing successful valid sensory data transmission.
		\item $\rho\in[0,1)$ is the discount factor, which indicates UAVs' evaluation of the rewards that obtained in the future (or in the past).
	\end{itemize}
\end{definition}

In the framework, the UAVs are informed of the rewards in the last cycle by the BS.
Specifically, we assume that the BS informs each UAV whether the sensory data transmitted in the previous cycle (if exists) is valid sensory data at the beginning of the next cycle.
For each UAV, it considers its reward in the $k$-th cycle to be $1$ if the BS informs that the valid sensory data has been received by the BS successfully at the beginning of the $(k+1)$-th cycle. 
\textcolor[rgb]{0,0,0}{The probability for UAV $i$ to obtain one reward after the cycle is equal to the probability for it to transmit valid sensory data to the BS successfully in the cycle, i.e., $p_{R,i}= p_{\mathrm{sTx},i}$. }
\textcolor[rgb]{0,0,0}{Since the probability of successful valid sensory data transmission is influenced by both the successful sensing probability and the successful} \textcolor[rgb]{0,0,0}{transmission probability, the UAV's trajectory learning process is associated with the sensing and transmission processes through the obtained reward in each cycle.}

Under the reinforcement learning framework for the UAV trajectory design, the following two kinds of reinforcement learning algorithms can be adopted, which are single-agent Q-learning algorithm and multi-agent Q-learning algorithm.

\subsubsection{Single-agent Q-learning Algorithm}
One of the most basic reinforcement learning algorithm is single-agent Q-learning algorithm \cite{Watkins1989Learning}. 
It is a form of model-free reinforcement learning and provides a simple way for the agent to learn how to act optimally.
The algorithm learns the optimal state-action value function $Q^*$, which then defines the optimal policy.
In its simplest form, the agent maintains a table containing its current estimates of $Q^*(s,a)$. 
It observes the current state $s$ and selects the action $a$ that maximizes $Q(s,a)$ with some exploration strategies.
Q-learning has been studied extensively in single-agent tasks where only one agent is acting alone in an unchanging environment.

In the UAV trajectory design problem, multiple UAVs take actions at the same time. 
When each UAV adopts the single-agent Q-learning algorithm, it assumes that the other agents are part of the environment. 
Therefore, in the UAV trajectory design problem, the single-agent Q-learning algorithm can be adopted as follows.
For UAV $i$, upon receiving a reward $R_i$ after the end of the cycle and observing the next state $s_i'$, it updates its table of Q-values according to the following rule, 
\begin{equation}\label{equ: update rule for single-agent}
	Q_i(s_i,a_i) \leftarrow Q_i(s_i,a_i) + \alpha (R_i+\rho \max_{a_i'\in \mathcal A(s_i)}Q_i(s_i,a_i)),
\end{equation}
where $\alpha\in(0,1)$ is the learning rate. 
With the help of \cite{bowling2003multiagent}, the single-agent Q-learning algorithm for UAV trajectory design of UAV $i$ can be summarized in Alg.~\ref{alg: single-agent q-learning algorithm for uav trajectory design}.

\begin{algorithm}[t]  
  \caption{Single-agent Q-learning Algorithm for UAV Trajectory Design of UAV $i$.}  
  \label{alg: single-agent q-learning algorithm for uav trajectory design}
  \begin{algorithmic}[1]  
    \REQUIRE  
    Learning ratio sequence ($\{\alpha_k\}\in(0,1]$); Exploration ratio ($\{\epsilon_k\}>0$);
    \STATE Initialize $Q_i(s_i,a_i) := 0$, $\forall s_i\in \mathcal S_p,~ a_i\in \mathcal A_i(s_i)$, $\pi_i(s_i,a_i) := \frac{1}{|A(s_i)|}$.
    \FOR{each cycle $k$}
    \STATE With probability $\epsilon^{(k)}$, choose action $a_i$ from the strategy at the state $\pi_i(s_i)$, or with probability $1-\epsilon^{(k)}$, randomly choose an available action for exploration;
    \STATE Perform the action $a_i$ in the $k$-th cycle;
    \STATE Observe the transited state $s_i'$ and the reward $R_i$;
    \STATE Select action $a_i'$ in the transited state $s_i'$ according to the strategy in state $s_i'$, i.e., $\pi_i(s_i')$;
    \STATE Update the Q-function for the former state-action pair, i.e., $Q_i(s_i,a_i):=Q_i(s_i,a_i) + \alpha_k(R_i+\rho Q(s_i',a_i')-Q_i(s_i,a_i) )$;
    \STATE Update the strategy at state $s_i$ as $\pi_i(s_i) := \argmax_{m} Q_i(s_i,m)$;
    \STATE Update the state $s_i := s_i'$ for the next cycle;
    \ENDFOR
    \end{algorithmic}  
\end{algorithm}   

\subsubsection{Multi-agent Q-learning Algorithm}
Although single-agent Q-learning algorithm has many favorable properties such as small state space and easy implementation, it lacks of consideration on the states and the strategic behaviors of other agents.
Therefore, we adopt a multi-agent Q-learning algorithm called opponent modeling Q-learning to solve the UAV trajectory design problem, which enables the agent to adapt to other agents' behaviors. 

Opponent modeling Q-learning is an effective multi-agent reinforcement learning algorithm \cite{uther1997adversarial,claus1998dynamics}, in which explicit models of the other agents are learned as stationary distributions over their actions.
These distributions, combined with learned joint state-action values from standard temporal differencing, are used to select an action in each cycle.

Specifically, at the beginning of the cycle, UAV $i$ selects an action $a_i$ to maximize the expected discounted reward according to the observed frequency distribution of other agents' action in the current state $\bm s$, i.e.,
\begin{equation} \label{equ: action selection in multi-agent alg.}
a_i = \pi_i(\bm s)= \argmax_{a_i''} \sum_{\sum a_{-i}''}	\frac{\Phi(\bm s, \bm a_{-i}'')}{n(\bm s)}Q_i(\bm s,(a_i'',\bm a_{-i}'')) 
\end{equation}
 in which the location profile  $\bm s = (s_1,...,s_N)$ observed by agent $i$ is adopted as state,
 $\pi_i(\bm s)$ denotes the strategy of UAV $i$ in state $\bm s$,
 $\Phi(\bm s, \bm a_{-i}'')$ denotes the number of times for the agents other than agent $i$ to select action profile $\bm a_{-i}''$ in the state $\bm s$, 
 and $n(\bm s)$ is the total number of times the state $\bm s$ has been visited.
 
 After the agent $i$ observes the transited state $\bm s'$, the action profile $( a_i, \bm a_{-i})$, and the reward in the previous cycle after performing the action $a_i$, it will update its table of Q-value as follows.
 \begin{equation} \label{equ: label update in multi-agent alg.}
Q_i(\bm s,( a_i,\bm a_{-i}))\!=\!(1-\alpha)Q_i(\bm s,( a_i,\bm a_{-i})) + \alpha(R_i+\rho V_i(\bm s')),
 \end{equation}
 in which $V_i(\bm s') = \max _{a_i''} \sum_{\bm a_{-i}''} \frac{\Phi(\bm s', \bm a_{-i}'')}{n(\bm s')} Q(s,( a_i'',\bm a_{-i}''))$ indicating that agent $i$ considers the action taken in the new state to maximize the expected discounted reward based on the empirical action profile distribution.
With the help of \cite{claus1998dynamics}, the multi-agent Q-learning algorithm for UAV trajectory design can be summarized in Alg.~\ref{alg: opponent modeling q-learning algorithm for uav trajectory design}.

\begin{algorithm}[t]  
  \caption{Opponent Modeling Q-learning Algorithm for UAV Trajectory Design of UAV $i$.}  
  \label{alg: opponent modeling q-learning algorithm for uav trajectory design}
  \begin{algorithmic}[1]  
    \REQUIRE  
    Learning ratio sequence ($\{\alpha^{(k)}\}\in(0,1]$); Exploration ratio sequence ($\{\epsilon^{(k)}\}>0$);
    \STATE Initialize $Q_i(\bm s, (a_i, \bm a_{-i})) := 0, \forall \bm s \in \prod_i^N \mathcal S_i,~ a_i\in \mathcal A_i(s_i), \bm a_{-i}\in \prod_{j\neq i}^N \mathcal A_j \pi_i(\bm s, a_i) := \frac{1}{|A(\bm s)|}$.
    \FOR{each cycle $k$}
    \STATE With probability $\epsilon^{(k)}$, choose action $a_i$ from the strategy at the state $\pi_i(\bm s)$, or with probability $1-\epsilon^{(k)}$, randomly choose an available action for exploration;
    \STATE Perform the action $a_i$ in the $k$-th cycle;
    \STATE Observe the transited state $\bm s'$ and the reward $R_i$;
    \STATE Select action $a_i'$ in the transited state $\bm s'$ according to the strategy in state $\bm s'$ according to (\ref{equ: action selection in multi-agent alg.});
    \STATE Update the Q-function for the former state-action pair according to (\ref{equ: label update in multi-agent alg.});
    \STATE Update the strategy at state $\bm s$ to the action that maximizes the expected discounted reward according to (\ref{equ: action selection in multi-agent alg.});
    \STATE Update the state $\bm s :=\bm s'$ for the next cycle;
    \ENDFOR
    \end{algorithmic}  
\end{algorithm}   

\subsection{Enhanced Multi-agent Q-learning Algorithm for UAV Trajectory Design}
In the opponent modeling multi-agent reinforcement learning algorithm, UAVs face need to tackle too many state-action pairs, resulting in a slow convergence speed.
Therefore, we enhance the opponent modeling Q-learning algorithm in the UAV trajectory design problem by reducing the available action set and adopting an model-based reward representation. 
These two enhancing approaches are elaborated as follows, and the proposed enhanced multi-UAV Q-learning algorithm is given in Alg.~\ref{alg: enhanced q-learning algorithm}.
\begin{algorithm}[t]  
  \caption{Enhanced Multi-UAV Q-learning Algorithm for Trajectory Design of UAV $i$.}  
  \label{alg: enhanced q-learning algorithm}
  \begin{algorithmic}[1]  
    \REQUIRE  
  Learning ratio sequence ($\{\alpha^{(k)}\}\in(0,1]$); Exploration ratio ($\{\epsilon^{(k)}\}>0$);
    \FOR{each cycle $k$}
    \STATE Obtain the available action set $\mathcal A^+_j(s_j),~\forall j\in[1,N]$ for the current state $\bm s$ according to Def. \ref{def: available action set}.
	\IF{state $\bm s$ has not been reached before}
	\STATE Initialize $Q_i(\bm s, \bm a) := p_{sTx,i}(\bm s,\bm a), \forall \bm s \in \prod_i^N \mathcal S_i,~ \bm a \in \prod_{j=1}^N \mathcal A_j^+(s_j), \pi_i(\bm s, a_i) := \frac{1}{|\mathcal A_i^+(\bm s)|}$.
	\ENDIF
    \STATE With probability $\epsilon^{(k)}$, choose action $a_i$ from the strategy at the state $\pi_i(\bm s)$, or with probability $1-\epsilon^{(k)}$, randomly choose an available action for exploration;
    \STATE Perform the action $a_i$ in the $k$-th cycle;
    \STATE Observe the transited state $\bm s'$ and the action profile $\bm a$ in the previous state;
    \STATE Select action $a_i'$ in the transited state $\bm s'$ according to the strategy in state $\bm s'$ according to (\ref{equ: action selection in multi-agent alg.});
    \STATE Calculate the successful valid sensory data transmission probability in the previous state transition $p^{(k)}_{sTx,i}(\bm s,\bm a)$ and consider it as the reward $\hat{R}_i$.
    \STATE Update the Q-function for the former state-action pair according to (\ref{equ: label update in multi-agent alg.}), substituting $R_i$ with $\hat{R}_i$;
    \STATE Update the strategy at state $\bm s$ to the action that maximizes the expected discounted reward according to (\ref{equ: action selection in multi-agent alg.});
    \STATE Update the state $\bm s := \bm s'$ for the next cycle;
    \ENDFOR
    \end{algorithmic}  
\end{algorithm}   

\subsubsection{Available Action Set Reduction}
\begin{figure}[!t] % Example image
	\center{\includegraphics[width=0.8 \linewidth] {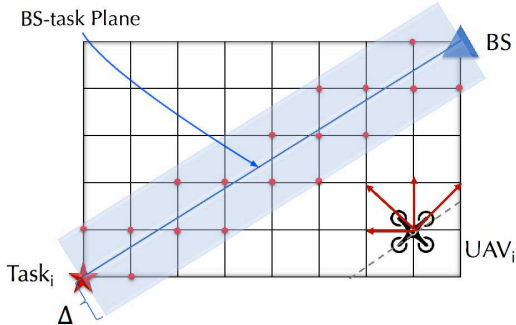}}
	\caption{Illustration on the constrained available action set of UAV $i$.}
	\label{fig: available action set}
		\vspace{-0.5em}
\end{figure}

It can be observed that although the UAVs are possible to reach all the location points in the finite location space $\mathcal S_p$, it makes no sense for the UAVs to move away from the vertical plane passing the BS and their tasks, i.e. the BS-task plane, which descreases the successful sensing probability as well as the successful transmitting probability.
Therefore, we confine the available action set of the UAV to the actions which does not increase the horizontal distance between it and the BS-task plane, which is shown in Fig.~\ref{fig: available action set} (the arrows).

Ideally, the UAVs should be in the BS-task plane and only move within the plane. 
However, since the location space is discrete, the UAV cannot only move within the BS-task plane in general, and needs to deviate from the plane in order to reach different locations near the plane. 
Therefore, we mitigate the constraint by allowing the UAV to move to the location from which the distance to the BS-task plane is within $\Delta$, as the spots shown in Fig.~\ref{fig: available action set}.
The reduced available action set of UAV $i$ at state $s_i= (x_i,y_i,h_i)$ can be defined as follows.

\begin{definition}[Reduced available action set of UAV $i$] \label{def: available action set}
	Suppose UAV $i$ is at the state $s_i = (x_i,y_i,h_i)$, denote the location of its task as $S_i =(X_i ,Y_i ,0)$, and denote the location of BS as $S_0=(0 ,0 ,H_0)$, the action $a=(a_x,a_y,a_h)$ in the reduced available action set $\mathcal A_i^+(s_i)$ satisfies the following conditions.
	\begin{enumerate}
		\item $\mathrm{Dist}(s_i+a;S_i,S_0) \leq \mathrm{Dist}(s_i;S_i,S_0)$ or $\mathrm{Dist}(s_i+a;S_i,S_0) \leq \Delta$;
		\item $x_i+a_x\in [\min(x_i,S_i,0),\max(x_i,S_i,0)]$,\\$y_i+a_y\in [\min(y_i,Y_i,0),\max(y_i,Y_i,0)]$, and \\ $h_i+a_h \in [h_{\min}, h_{\max}]$.
	\end{enumerate}
Here $\mathrm{Dist}(s;S_i,S_0)$ denotes the horizontal distance between the location $s$ to the vertical plane passing through $S_i$ and $S_0$. 
\end{definition}

In Def. \ref{def: available action set}, condition 1) limits the actions to those leading the UAV to a location near the BS-task plane, and condition 2) stops the UAV from moving away from the cross region between the location of its task and the BS.

%in the above definition, we neglect the condition that the action does not lead the UAV away from the rectangular area defined by the max min horizontal location among the task-UAV-BS.

Moreover, instead of initializing Q-function for all the possible state-action pair at the beginning, we propose that the UAVs initialize the Q-function only when the state is actually reached, and the actions are in the reduced available action set of the current state.
In this way, the state sets of UAVs are reduced to some smaller sets, which makes the reinforcement learning more effective and converge faster.

\subsubsection{Model-based Reward Representation}
In both the single-agent Q-learning algorithm and the opponent modeling Q-learning algorithm, the UAVs update their Q-values based on the information provided by the BS, which indicates the validity of the latest transmitted sensory data. 
Nevertheless, since the UAVs can only observe the reward to be either 1 or 0, the Q-functions converge slowly and the performance of the algorithms is likely to be poor.

Therefore, in this paper, we propose the UAVs update their Q-functions based on the probability of successful valid sensory data transmission obtained in Section \ref{sec: analysis on reward}. 
In other words, UAV $i$ calculates the probability $p_{\mathrm{sTx},i}$ after observing the state-action profile $(\bm s, ( a_i, \bm a_{-i}))$ in the previous cycle according to (\ref{equ: successful valid sensory data transmission prob.}), and considers it as the reward $R_i$ for the $k$-th cycle.

Moreover, to make the reinforcement learning algorithm converge more quickly, in the initialization of the enhanced multi-UAV Q-learning algorithm, we propose that UAV $i$ initializes its $Q_i(\bm s,(a_i,\bm a_{-i}))$ with the calculated $p_{sTx,i}$ for the state-action pair.
In this way, the update of the Q-function is more accurate and the reinforcement learning algorithm is expected to have higher convergence speed.

\textcolor[rgb]{0,0,0}{
\emph{\textbf{Remark} (Signaling in UAVs' learning algorithms)}
In the above mentioned reinforcement learning algorithms, UAVs need to know the locations of themselves in the beginning of each cycle, and the rewards in the last cycle associated with their actions taken. 
Besides, for multi-agent Q-learning algorithm and the proposed enhanced multi-UAV Q-learning algorithm, UAVs also need to know the  locations of other UAVs before determine their flying directions in each cycle.
This information gathering can be done in beaconing phase of the cycle as described in Section \ref{sec: sense-and-send cycle}, in which the BS can include the rewards of UAVs in the last cycle in the broadcasting frame.
}

\subsection{\textcolor[rgb]{0,0,0}{Analysis of Reinforcement Learning Algorithms}}
In the final part of this section, we analyze the convergence, the complexity, and the scalability of the proposed reinforcement learning algorithms.
\subsubsection{Convergence Analysis}
For the convergence of the reinforcement learning algorithms, it has been proved in \cite{jaakkola1994convergence} that under certain conditions, single agent Q-learning algorithm is guaranteed to converge to the optimal $Q^*$. 
In consequence, the policy $\pi$ of the agent converges to the optimal policy $\pi^*$.
It can be summarized in the following Theorem \ref{theo: convergence of Q}.
\begin{theorem} \label{theo: convergence of Q}
\emph{(Convergence of Q-learning Algorithm)}
The Q-learning algorithm given by 
\begin{align}
Q^{(k+1)}(s^{(k)},&a^{(k)}) = \big(1-\alpha^{(k)}\big)Q^{(k)}(s^{(k)},a^{(k)}) \\
&+ \alpha^{(k)}[R(s^{(k)},a^{(k)}) + \gamma \max_{a'}Q(s^{(k+1)},a')] \nonumber
\end{align}
converges to the optimal $Q^*$ values if
\begin{enumerate}
\item The state and action spaces are finite.
\item $\sum_k\alpha^{(k)} = \infty$ and $\sum_k (\alpha^{(k)})^2<\infty$.
\item The variance of $R(s,a)$ is bounded.
\end{enumerate}
\end{theorem}

%The proof of Theorem \ref{theo: convergence of Q} can be found in \cite{jaakkola1994convergence}. 
Therefore, in the multi-agent reinforcement learning cases, if other agents play, or converge to stationary strategies, the single-agent reinforcement learning algorithm also converges to an optimal response.

However, it is generally hard to prove convergence with other players that are simultaneously learning. 
This is because that when agent is learning the value of its actions in the presence of other agents, it is a non-stationary environment. 
Thus, the convergence of Q-values is not guaranteed.
The theoretical convergence of the Q-learning in multi-agent cases are guaranteed only in few situations such as iterated dominance solvable games and team games \cite{bowling2003multiagent}.
Like single-agent Q-learning algorithm, the convergence of opponent modeling Q-learning is not generally  guaranteed, except for in the setting of iterated dominance solvable games and team matrix game \cite{claus1998dynamics}.

Therefore, in this paper, we adopt $\alpha^{(k)}= 1/k^{2/3}$ in \cite{singh2000nash} which satisfies the conditions for convergent in single-agent Q-learning, and analyze the convergence of the reinforcement learning in the multi-agent case through simulation results which will be provided in Section \ref{sec: simulation result}.

\subsubsection{Complexity Analysis}
For the single-agent Q-learning algorithm, the computational complexity in each iteration is $\mathcal O(1)$, since the UAV does not consider the other UAVs in the learning process.
For the multi-agent Q-learning algorithm, the computational complexity in each iteration is $\mathcal O(2^N)$, due to the calculation of the expected discounted reward in (\ref{equ: action selection in multi-agent alg.}).

As for the proposed enhanced multi-UAV Q-learning algorithm, each UAV needs to calculate the probability for successful valid data transmission based on Alg.~\ref{alg: solve successful transmission probability}. 
It can be seen that the recursive Alg.~\ref{alg: solve successful transmission probability} runs for at most $2^{CT_u}$ times and is of complexity $\mathcal O(N)$, which is smaller than $\mathcal O(2^N)$. 
Therefore the complexity of the proposed enhanced algorithm is also $\mathcal O(2^N)$, due to the expectation over the joint action space.

Although the computational complexity of the enhanced multi-UAV Q-learning algorithm in each iteration is in the same order with opponent modeling Q-learning algorithm, it reduces the computational complexity significantly and speeds up the convergence by the following means.

\begin{enumerate}[(1)]
	\item Due to the available action set reduction, the available action set of each UAV is at least reduced to one-half its original size. This makes the joint action space to be $2^N$ times smaller. 
	\item The reduced available action set leads to a much smaller state space of each UAV. 
For example, for UAV $i$ and its task at $(X_i,Y_i,0)$, the original size of its state space can be estimated as $\pi R_{\max}^2(h_{\max}-h_{\min})/\Delta^3$, and the size of its state space after available action set reduction is $2(X_i+Y_i)(h_{\max}-h_{\min})/\Delta^2$, which is $2\Delta/(\pi R_{\max})$ smaller than the original one.
\item The proposed algorithm adopts model-based reward representation, which makes the Q-value updating in the enhanced multi-UAV Q-learning algorithm to be more precise, and saves the number of iterations needed to estimate the accurate Q-values of the state-action pairs.
\end{enumerate}

\subsubsection{Scalability Analysis}
%Therefore, in this paper we adopt $\alpha_k = 1/k^{2/3}$ as in \cite{Singh:2000:NCG:2073946.2074009}, which satisfies the condition (\ref{equ: converge condition}), in order to ensure that the reinforcement learning algorithms converge for single agent reinforcement learning. 
With the growth of the number of UAVs, the state spaces of UAVs in the multi-agent Q-learning algorithm and the enhanced multi-UAV Q-learning algorithm grow exponentially.
Besides, it can be seen that the enhanced multi-UAV Q-learning algorithm still has exponential computational complexity in each iteration, and thus, it is not suitable for large-scale UAV networks.
%This implies that the required explorations and duration for UAVs to estimate the Q-values of state-action pairs also grow exponentially.

To adapt the algorithms for large-scale UAV networks, the reinforcement learning methods need to be combined with function approximation in order to estimate Q-values efficiently.
The function approximation takes examples from a desired function, Q-function in the case of reinforcement learning, and generalizes from them to construct an approximation of the entire function.
In this regard, it can be used to efficiently estimate the Q-values of the state-action pairs in the entire state space when the state space is large.

%Besides, to avoid the high-complexity in computing the expectation of the Q-value associated with each action, we can leave it to be learnt implicitly instead of explicitly calculating as in the opponent-modeling algorithm. 
%Although in this way we loss some accuracy in modeling the action selection of other agents, the computational complexity in each iteration can be reduced to $\mathcal O(n)$.
%Therefore, combing with the function approximation, we can make the multi-agent reinforcement learning algorithms applicable to large-scale networks.

%æ­¤å¤ï¼æä»¬è¿è¦æ¾å¼opponent modeling éé¢å¯¹äºåå¼çè®¡ç®ï¼è¿ä¸ªæ¾å¼çä¾æ®å¯ä»¥åèããã

\section{Simulation Results}\label{sec: simulation result}
In order to evaluate the performance of the proposed reinforcement learning algorithms for the UAV trajectory design problem, simulation results are presented in this section.
Specifically, we use MATLAB to build a frame-level simulation of the UAV sense-and-send protocol, based on the system model described in Section \ref{sec: system model} and the parameters in Tab. \ref{tab: simulation parameter}. 
Besides, the learning ratio in the algorithm is set to be $\alpha^{(k)} = 1/k^{2/3}$ in order to satisfy the converge condition in Theorem 1. 
The exploration ratio is set to be $\epsilon^{(k)} = 0.8e^{-0.03k}$, which approaches 0 when $k\rightarrow \infty$.

\begin{table}[!t]
\caption{Simulation Parameters}
\centering
\begin{tabular}{| c | c |}
\Xhline{1.pt}
\textbf{Parameter}	&	\textbf{Value}\\
\hline
\hline
BS height $H$ & 25 m \\
\hline
Number of UAVs $N$ & 3 \\
\hline
Noise power $N_0$ & -85 dBm \\
\hline
BS decoding threshold $\gamma_{th}$ & 10 dB\\
\hline
UAV sensing parameter $\lambda$ & $10^{-3}$/s\\
\hline
UAV transmit power $P_u$ & 10 dBm\\
\hline
Duration of frame $t_f$ & 0.1 s\\
\hline
Distance between adjacent spatial points $\Delta$ & 25 m\\
\hline
UAVs' minimum flying height $h_{\min}$ & 50 m\\
\hline
UAVs' maximum flying height $h_{\max}$ & 150 m\\
\hline
Discounted ratio $\rho$ & 0.9\\
\hline
Duration of beaconing phase in frames $T_b$ & 3\\
\hline
Duration of sensing phase in frames $T_s$ & 5\\
\hline
Duration of transmission phase in frames $T_u$ & 5\\
\Xhline{1.2pt}
\end{tabular}
\label{tab: simulation parameter}
\end{table}

\begin{figure}[!t] % Example image
	\center{\includegraphics[width=1\linewidth] {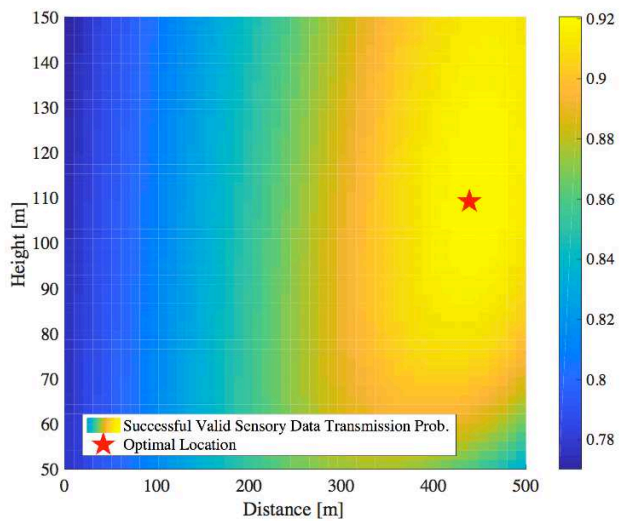}}
	\caption{Successful valid sensory data transmission probability versus the location in the task-BS surface.}
	\label{fig: suc. val. sense data tx. prob.}
		\vspace{-0.5em}
\end{figure}

Fig.~\ref{fig: suc. val. sense data tx. prob.} shows UAV 1's successful valid sensory data transmission probability versus UAV 1's height and its distance to the BS, given that the other two UAVs are located at their initial locations, task 1 is located at $(500,0,0)$, and the locations of UAV 2 and UAV 3 are fixed at $(-125,125,75), (-125,-125,75)$, respectively.
It can be seen that the optimal point at which UAV 1 has the maximum successful valid sensory data transmission probability is located in the region between BS and task 1.
This is because when the UAV approaches the BS (task), the successful sensing (transmission) probability drops since it moves far away from the task (BS).
Besides, it can be seen from the transmission model in Section \ref{sec: UAV transmission} that when the height of the UAV increases, the LoS probability for the transmission channel will increase, and thus, the successful uplink transmission probability of the UAV increases. 
Therefore, the optimal point for UAV 1 to sense-and-send is above rather than on the BS-task line, where UAV 1 can be closer to both the BS and its task.

\begin{figure}[!t] % Example image
	\center{\includegraphics[width=0.9\linewidth] {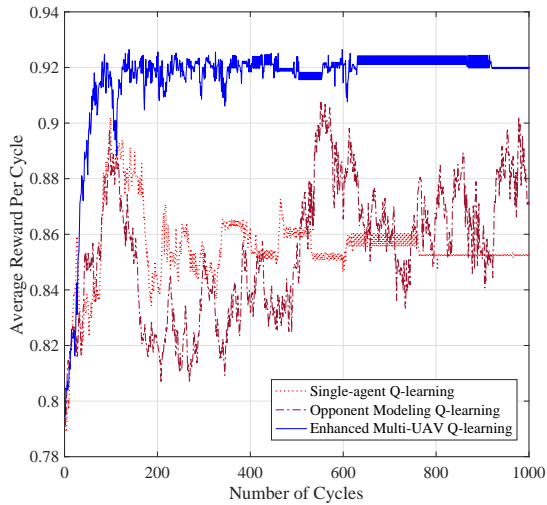}}
	\caption{UAVs' average reward per cycle versus number of cycles of different reinforcement learning algorithms.}
	\label{fig: per cyc. reward vs. cyc.}
	\vspace{-1em}
\end{figure}

\begin{figure}[!t] % Example image
	\center{\includegraphics[width=0.9\linewidth] {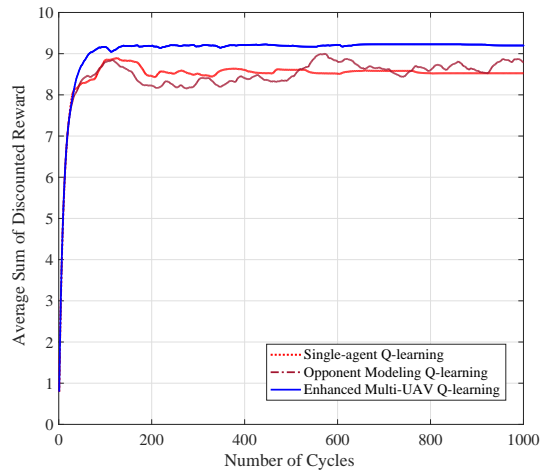}}
	\caption{UAVs' average discounted reward versus number of cycles of different reinforcement learning algorithms.}
	\label{fig: disc. reward vs. cyc.}
		\vspace{-0.5em}
\end{figure}

Fig.~\ref{fig: per cyc. reward vs. cyc.} and Fig.~\ref{fig: disc. reward vs. cyc.} show the average reward per cycle and the average total discounted reward of the UAVs versus the number of cycles in different reinforcement learning algorithm, in which tasks 1, 2 and 3 are located at $(500,0,0),(-250\sqrt{2},250\sqrt{2},0)$ and $(-250\sqrt{2},-250\sqrt{2},0)$, respectively. 
It can be seen that compared to the single-agent Q-learning algorithm, the proposed algorithm converges to a higher average reward for the UAVs. 
This is because the enhanced multi-UAV Q-learning algorithm takes the states of all the UAVs into consideration, which makes the estimation for Q-function of each UAV more precise. 
Besides, it can also be seen that compared to the opponent modeling Q-learning algorithm, the proposed algorithm converges faster, due to the available action set reduction and the reward representation.

\begin{figure}[!t] % Example image
	\center{\includegraphics[width=0.9\linewidth] {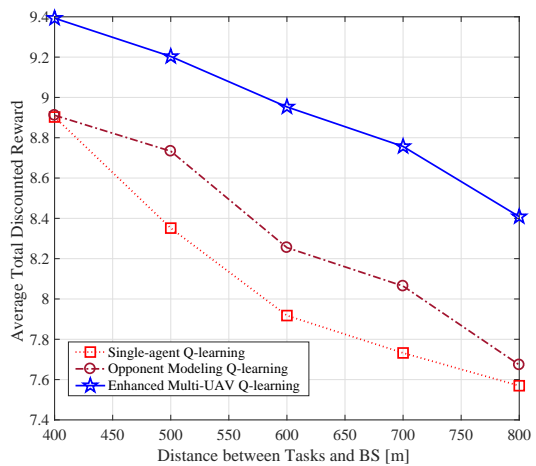}}
		\vspace{-1em}
	\caption{UAVs' average discounted reward versus distance between tasks and BS in different reinforcement learning algorithms.}
	\label{fig: per cyc. reward vs. task dist.}
	\vspace{-0.5em}
\end{figure}

Moreover, in Fig.~\ref{fig: per cyc. reward vs. task dist.}, we can observe that for different distances between the tasks and the BS, the proposed algorithm converges to a higher average discounted reward for UAVs after 1000 cycles compared to two other algorithms. 
It can be seen that the average discounted reward in the algorithms decreases with the increment of the distance between the BS and the tasks.
Nevertheless, the decrement in the proposed algorithm is less than those in the other algorithms.
This indicates that the proposed algorithm is more robust to the variance of the tasks' location.

 \begin{figure}[!t] % Example image
	\center{\includegraphics[width=0.9\linewidth] {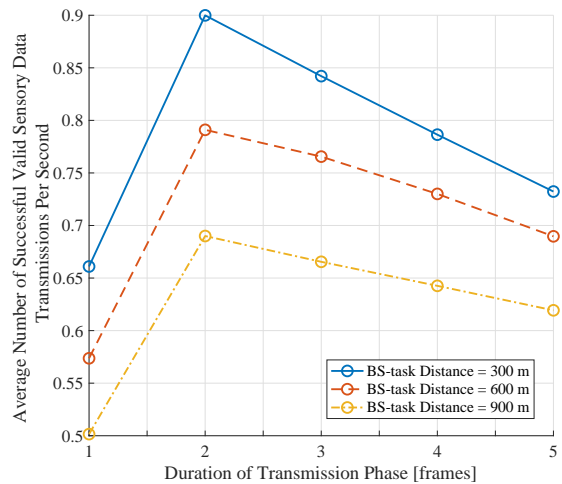}}
	\caption{Average number of successful valid sensory data transmissions per second versus duration of transmission phase $T_u$ under different task distance conditions.}
	\vspace{-0.5em}
	\label{fig: disc. reward vs. cycle under diff. LR and ER case}
\end{figure}

 \begin{figure}[!t] % Example image
	\center{\includegraphics[width=0.9\linewidth] {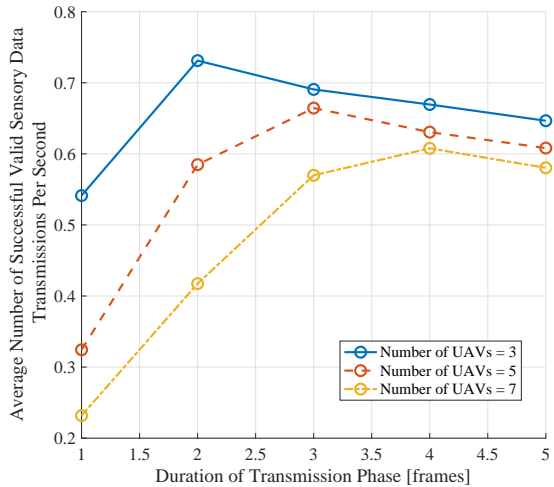}}
	\caption{\textcolor[rgb]{0,0,0}{Average number of successful valid sensory data transmissions per second versus duration of transmission phase $T_u$ under different number of UAVs. Distance between the BS and the tasks $=800$ m.}}
	\vspace{-0.5em}
	\label{fig: average valid sensory data transmitted per second}
\end{figure}

Fig.~\ref{fig: disc. reward vs. cycle under diff. LR and ER case} shows the average number of successful valid sensory data transmissions per second of the proposed algorithm versus the duration of the transmission phase $T_u$, under different conditions of the distance between the tasks and the BS.  
It can be seen that the average number of successful valid sensory data transmissions per second first increases and then decreases with the increment of $T_u$. 
When $T_u$ is small, the successful uplink transmission probability increases rapidly with the increment of $T_u$. 
However, when $T_u$ is large, the successful uplink transmission probability is already high and increases slightly when $T_u$ becomes larger. 
Therefore, the average number of successful valid sensory data transmissions per second drops due to the increment of cycles' duration.

\textcolor[rgb]{0,0,0}{Fig.~\ref{fig: average valid sensory data transmitted per second} shows the average number of successful valid sensory data transmissions per second versus $T_u$ with different number of UAVs. 
It can be seen that when the number of UAVs increases, the average number of successful valid sensory data transmissions per second decreases. 
This is because the competition among the UAVs for the limited SCs becomes more intensive.
Besides, when the number of UAVs increases, the optimal duration of the transmission phase becomes longer.
This indicates that the BS needs to choose the optimal $T_u$ according to the number of UAVs in order to improve the spectrum efficiency.
}
\section{Conclusion}
In this paper, we have adopted the reinforcement learning framework to solve the trajectory design problem in a decentralized manner for the UAV to perform different real-time sensing task.
We have proposed a sense-and-send protocol to coordinate multiple UAVs performing real-time sensing tasks.
To evaluate the performance of the protocol, we have proposed a recursive algorithm to solve the successful valid sensory data transmission probability in the protocol.
Besides, under the reinforcement learning framework, we have proposed an enhanced multi-UAV Q-learning algorithm to solve the decentralized trajectory problem.
The simulation results showed that the proposed algorithm converges faster and achieves higher rewards for the UAVs. 
It was also shown in simulation that our proposed algorithm was more robust to the increment of tasks' distance, comparing to single-agent and opponent modeling Q-learning algorithms.
Moreover, the simulation also showed that the BS needs to increase the duration of the transmission phase to improve the spectrum efficiency when the number of UAVs increases.
%%%%%%%%%%%%%%%%%%%%%%%%%%%%%%%%%%%%%%%%%%

%%%%%%%%%%%%%%%%%%%%%%%%%%%%%%%%%%%%%%%%%%
\begin{appendices}
\section{Proof of Proposition \ref{prop: theo res}}\label{appx: a}
Denoting the UAVs' probability for successful uplink transmission as $p_u$ and their probability for successful sensing as $p_s$, the average number of valid sensory data transmissions per second can be calculated as 
\begin{equation*} \label{equ: theo res}
N_{vd} =N\cdot \frac{p_s(1-(1-p_u)^{\frac{CT_u}{N}})}{(T_b+T_s+T_u)t_f},
\end{equation*}
in which $t_f$ is the duration of single frame in seconds.

The partial derivative of $N_{vd}$ with respect to $T_u$ can be calculated as 
\begin{equation*}
	\frac{\partial N_{vd}}{\partial T_u} = \frac{p_sF(T_u) }{t_f (T_b+T_s+T_u)^2}
\end{equation*}
in which $F(T_u) = p_f^{\frac{CT_u}{N}}(N-C(T_b+T_s+T_u)\ln p_f)-N$, and $p_f = 1-p_u$. 
Taking partial derivative of $F(T_u)$ with regard to $T_u$, it can be derived that $\partial F(T_u)/\partial T_u = -C^2p_f^{CT_u/N}(T_s+T_b+T_u)\ln p_f/N<0$. 
Besides, when $T_u \rightarrow \infty$, $F(T_u) \rightarrow -N$ and $N_{vd}\rightarrow 0$, and when $T_u = 0$, $N_{vd}=0$.
Therefore, $\partial F(T_u)/\partial T_u<0$ indicates that there is a unique maximum point for $N_{vd}$ when $T_u \in(0,\infty)$.

The maximum of $N_{vd}$ is reached when $F(T_u^*) = 0$, in which $T_u^*$ can be solved as 
\begin{equation*}
T_u^* =\frac{N}{C\ln p_f}(1+W_{-1}(-\frac{p_f^{\frac{CT_u}{N}}}{e}))-T_b-T_s,
\end{equation*}
where $W_{-1}(\cdot)$ denotes the lower branch of Lambert-W function \cite{Corless1996}.$\hfill\blacksquare$
\end{appendices}


\begin{thebibliography}{20}
\bibitem{wang2017taking}
J. Wang, C. Jiang, Z. Han, Y. Ren, R. G. Maunder, and L. Hanzo, ``Taking drones to the next level: Cooperative distributed unmanned-aerial-vehicular networks for small and mini drones," \emph{IEEE Veh. Technol. Mag.}, vol. 12, no. 3, pp. 73–82, Jul. 2017.
\bibitem{puri2007statistical}
A. Puri, K. Valavanis, and M. Kontitsis, ``Statistical profile generation for traffic monitoring using real-time UAV based video data," in \emph{Mediterranean Conf. Control \& Automation.} Athens, Greece, Jun. 2007.
\bibitem{alsalam2017autonomous}
B. H. Y. Alsalam, K. Morton, D. Campbell, and F. Gonzalez, ``Autonomous UAV with vision based on-board decision making for remote sensing and precision agriculture," in \emph{Aerosp. Conf.} Big Sky, MT, USA, Mar. 2017.
\bibitem{casbeer2006cooperative}
D. W. Casbeer, D. B. Kingston, R. W. Beard, and T. W. McLain, ``Cooperative forest ﬁre surveillance using a team of small unmanned air vehicles," \emph{Int. J. Syst. Sci.}, vol. 37, no. 6, pp. 351–360, Feb. 2006.
\bibitem{van2016lte}
B. Van der Bergh, A. Chiumento, and S. Pollin, ``LTE in the sky: Trading off propagation benefits with interference costs for aerial nodes," \emph{IEEE Commun. Mag.}, vol. 54, no. 5, pp. 44–50, May 2016.


\bibitem{zhang2018cellular2}
S. Zhang, H. Zhang, B. Di, and L. Song, ``Cellular UAV-to-X communications: Design and optimization for multi-UAV networks," \emph{arXiv preprint arXiv:1801.05000}, Jan. 2018.
\bibitem{thammawichai2018optimizing}
M. Thammawichai, S. P. Baliyarasimhuni, E. C. Kerrigan, and J. B. Sousa, ``Optimizing communication and computation for multi-UAV information gathering applications," \emph{IEEE Trans. Aerosp. Electron. Syst.}, vol. 54, no. 2, pp. 601–615, Oct. 2018.
\bibitem{tisdale2009autonomous}
J. Tisdale, Z. Kim, and J. K. Hedrick, ``Autonomous UAV path planning and estimation," \emph{IEEE Robot. Autom. Mag.}, vol. 16, no. 2, Jun. 2009.
\bibitem{maza2009multi}
I. Maza, K. Kondak, M. Bernard, and A. Ollero, ``Multi-UAV cooperation and control for load transportation and deployment," \emph{J. Intell. \& Robot. Syst.}, vol. 57, no. 1-4, p. 417, Jan. 2010.
\bibitem{gu2006optimal}
G. Gu, P. Chandler, C. Schumacher, A. Sparks, and M. Pachter, ``Optimal cooperative sensing using a team of UAVs," \emph{IEEE Trans. Aerosp. Electron. Syst.}, vol. 42, no. 4, pp. 1446–1458, Oct. 2006.

\bibitem{yang2018real}
Y. Yang, Z. Zheng, K. Bian, L. Song, and Z. Han, ``Real-time proﬁling of fine-grained air quality index distribution using UAV sensing," \emph{IEEE Internet Things J.}, vol. 5, no. 1, pp. 186–198, Nov. 2018.
\bibitem{zhang2018joint}
S. Zhang, H. Zhang, Q. He, K. Bian, and L. Song, ``Joint trajectory and power optimization for UAV relay networks," \emph{IEEE Commun. Lett.}, vol. 22, no. 1, pp. 161–164, 2018.
\bibitem{bor2016efficient}
R. I. Bor-Yaliniz, A. El-Keyi, and H. Yanikomeroglu, ``Efficient 3-d placement of an aerial base station in next generation cellular networks," in \emph{IEEE ICC.} Kuala Lumpur, Malaysia, May 2016.
\bibitem{zhang2018cellular}
S. Zhang, H. Zhang, B. Di, and L. Song, ``Cellular controlled cooperative unmanned aerial vehicle networks with sense-and-send protocol," \emph{arXiv preprint arXiv:1805.11779}, May 2018.
\bibitem{Shakhov2017Experiment}
V. V. Shakhov and I. Koo, ``Experiment design for parameter estimation in probabilistic sensing models," \emph{IEEE Sensors J.}, vol. 17, no. 24, pp. 8431–8437, Oct. 2017.

\bibitem{Chakraborty2013Network}
A. Chakraborty, R. R. Rout, A. Chakrabarti, and S. K. Ghosh, ``On network lifetime expectancy with realistic sensing and traffic generation model in wireless sensor networks," \emph{IEEE Sensors J.}, vol. 13, no. 7, pp. 2771–2779, Apr. 2013.
\bibitem{3GPP2017R14}
3GPP TR 38.901, ``Study on channel model for frequencies from 0.5 to 100 GHz," Release 14, Dec. 2017.
\bibitem{3GPP2017R15}
3GPP TR 36.777, ``Enhanced LTE support for aerial vehicles," Release 15, Dec. 2017.
\bibitem{rice1944mathematical}
S. O. Rice, ``Mathematical analysis of random noise," \emph{Bell Syst. Tech. J.}, vol. 23, no. 3, pp. 282–332, Jul. 1944.
\bibitem{marcum1950table}
J. I. Marcum, ``Table of Q functions," Tech. Rep., U.S. Air Force Project RAND Res. Memo. M-339, ASTIA Document AD 1165451, Rand Corporation, Santa Monica, CA, Jan.1950.


\bibitem{Demange1986Multi}
G. Demange, D. Gale, and M. Sotomayor, ``Multi-item auctions," \emph{J. Political Economy}, vol. 94, no. 4, pp. 863–872, Aug. 1986.
\bibitem{Corless1996}
R. M. Corless, G. H. Gonnet, D. E. G. Hare, D. J. Jeffrey, and D. E. Knuth, ``On the Lambert W function," \emph{Advances in Comput. Math.}, vol. 5, no. 1, no. 1, pp. 329–359, Dec. 1996. [Online]. Available: https://doi.org/10.1007/BF02124750
\bibitem{Yang2004Multiagent}
E. Yang and D. Gu, ``Multiagent reinforcement learning for multi-robot systems: A survey," \emph{Dept. Comput. Sci. Univ. Essex}, 2004.
\bibitem{Watkins1989Learning}
C. J. C. H. Watkins, ``Learning from delayed rewards," Ph.D. dissertation, King’s College, Cambridge, 1989.
\bibitem{bowling2003multiagent}
M. Bowling, ``Multiagent learning in the presence of agents with limitations," Ph.D. dissertation, Dept. Comput. Sci., Carnegie Mellon Univ., Pittsburgh, PA, Tech. Rep., May 2003.

\bibitem{uther1997adversarial}
W. Uther and M. Veloso, ``Adversarial reinforcement learning," School Comput. Sci., Carnegie Mellon Univ., Pittsburgh, PA, [Onlilne]. Available: http://www.cs.cmu.edu/afs/cs/user/will/ www/papers/Uther97a.ps, Tech. Rep., 1997.
\bibitem{claus1998dynamics}
C. Claus and C. Boutilier, ``The dynamics of reinforcement learning in cooperative multiagent systems," in \emph{Conf. Innov. Appl. Artif. Intell}. Madison, WI, Jul. 1998.
\bibitem{jaakkola1994convergence}
T. Jaakkola, M. I. Jordan, and S. P. Singh, ``On the convergence of stochastic iterative dynamic programming algorithms," \emph{Neural Comput.}, vol. 6, no. 6, pp. 1185–1201, Nov. 1994. [Online]. Available: http://dx.doi.org/10.1162/neco.1994.6.6.1185
\bibitem{singh2000nash}
S. Singh, M. Kearns, and Y. Mansour, ``Nash convergence of gradient dynamics in general-sum games," in \emph{Proc. Conf. Uncertainty in Artificial Intell.}, San Francisco, CA, USA, Jul. 2000.
\end{thebibliography}
\end{document}